\documentclass{acm_proc_article-sp}
\usepackage[T1]{fontenc}
\usepackage{textcomp}
\usepackage{upquote}
\usepackage{listings}
\begin{document}

\title{MPIgnite: An MPI-Like Language and Prototype Implementation for Apache Spark}

\numberofauthors{2} 
\author{
\alignauthor
Brandon L.\@ Morris,~~~~ Anthony Skjellum\titlenote{Corresponding author}\\
       \affaddr{Auburn University}\\
       \affaddr{Dept.\@ of Computer Science and Software Engineering}\\
       \email{\{blm0026,skjellum\}@auburn.edu}
}


\maketitle
\begin{abstract}

Scale-out parallel processing based on MPI is a 25-year-old standard with at least another
decade of preceding history of enabling technologies  in the High Performance Computing community.  Newer frameworks such
as MapReduce, Hadoop, and Spark represent industrial scalable computing solutions that have
received broad adoption because of their comparative simplicity of use, applicability to relevant problems,
and ability to harness scalable, distributed resources.  While MPI provides performance and
portability, it lacks in productivity and fault tolerance.  Likewise, Spark is a specific
example of a current-generation MapReduce and data-parallel computing infrastructure that
addresses those goals but in turn lacks peer communication support to allow featherweight, highly
scalable peer-to-peer data-parallel code sections. 

The key contribution of this paper is to
demonstrate how to introduce the collective and point-to-point peer 
communication concepts of MPI into a Spark environment.  This is 
done in order to produce
performance-portable, peer-oriented and group-oriented communication services while
retaining the essential, desirable properties of Spark. Additional concepts of
fault tolerance and productivity are considered. This approach is offered in contrast to
adding MapReduce framework as upper-middleware based on a traditional MPI implementation
as baseline infrastructure.  
\end{abstract}



\keywords{MPI, Spark, data-parallel, task-parallel, Scala, peer-to-peer communication,
parallel closures} 

\section{Introduction}
Since its introduction, MPI has served as a standardized means to
communicate between processes peer- and group-wise in large-scale parallel applications.
The arrival of Inter\-net-scale data and the concomitant introduction of new
computational problems requiring enormous processing capabilities have
thrust the issues of High Performance Computing (HPC) into the spotlight and made them more
relevant than ever. Interestingly, the usage and interest of MPI
itself has not followed the ``data deluge trend''
\cite{hpc.dying.blog}. Instead, a myriad of open, high-level
frameworks have been created to address problems heretofore restricted to
the domain of traditional HPC. However, they are targeted at developers without the domain
expertise or access to expensive, specialized cluster or supercomputer infrastructure.

As large scale data analytics has become more mainstream, alternatives to
traditional HPC have developed rapidly, with novel solutions to the particular
challenges in the problem domains for which they were created. Many of the
most popular of these frameworks, such as Hadoop \cite{Shvachko2010} and the MapReduce
\cite{Dean2004}
family of implementations, focus on exploiting the data parallelism
that exists within the particular task (without intermediate communication between tasks
as in MPI). These solutions are generally
incongruous with MPI, where the lowest level of abstraction resides
in a message moving between a pair of peer processes, or else a collective
operation such as a numerical reduction ({\em e.g.}, parallel sum) over a group of processes.

Inspired by the capabilities and appeal of these high-level analytics
applications, we sought to introduce certain key concepts
of traditional HPC, particularly the notion of peer message passing, into 
a commercially driven HPC-like framework. Both the traditional HPC and modern analytics
communities would benefit greatly from such an amalgamation with symbiotic sharing of
low-level performance concerns with high level application
development. Additionally, the fault-tolerance advantages of the baseline
framework could offer distinct advantages to the overall application; features
not easy to achieve in current ``pure MPI'' middleware implementations and applications.

In this communication, we present a novel adaptation of the popular Apache Spark 
\cite{zaharia2010spark} data
processing engine that has been augmented to incorporate a message
passing Application Programmer Interface (API) for its parallel tasks. We  show that this relatively minor
modification to Spark gives rise to an original programming environment in which
task parallelism and data parallelism can be used interchangeably.
Furthermore, many of the recent advancements in cloud computing and
modern object-oriented software development can be leveraged freely,
all the while maintaining the core computing patterns and concepts of the original MPI standard.
In this effort, we have worked to retain these concepts and the semantics of
the MPI operations but have not emphasized  producing a literal language binding
of the MPI standard or subset thereof.  

While this work is still in its early stages, our design thus far,
as described here, will lay the groundwork for a marriage of two thought-to-be
mutually exclusive models of large scale distributed analytics and
data processing.  This combination
will enable data scientists  to utilize long-studied
algorithms with message passing based task-parallel solutions easily; conversely,
``traditional MPI developers'' will be able to prototype or develop  on a mature,
modern, high-level platform with full access to state-of-the-art
cluster data analysis\footnote{The integration  of literal MPI third-party libraries
in our system is reserved for future work.}.

The rest of the paper is organized as
follows. We offer a brief background in Section~\ref{sec:background} on several
recent advancements in high level HPC, with a particular focus on the Apache Spark project,
and recent work developing and integrating these models on existing
HPC infrastructures. In section~\ref{sec:approach}, we describe our approach for augmenting
the Spark framework itself to support message passing. Section~\ref{sec:examples}
describes certain examples ``in action.'' We discuss our
results in Section~\ref{sec:results}, and propose certain avenues for future work as well 
while offering conclusions in
Sections~\ref{sec:future-work}, and \ref{sec:conclusion}, respectively.

\section{Background}
\label{sec:background}

As essential background, we provide an overview of some
high level computing frameworks developed in recent years. We then give
an overview of the Apache Spark framework in particular, since it is
the focus of our work. We conclude with certain related work of implementing
Spark as middleware on HPC clusters.

\subsection{High-level, large-scale computing}

In this section, we consider ``non-traditional''  HPC environments and patterns
of programming,  originating from both industry and academia.

\subsubsection{MapReduce}

One of the most well-known and influential frameworks for large-scale,
distributed analytics and data processing is Map\-Reduce \cite{Dean2004},.
In MapReduce, data is distributed across a number of worker nodes by a master node,
operations are performed to transform ({\em i.e.,} map) the data to outputs,
and results are reduced to a final output. The simplicity of the
model, combined with its performance, led to its popularity and
adoption by data-hungry organizations such as Google.

A greater portion of the appeal of MapReduce models apparently lies in their simplicity: many of the 
difficulties (whether actual or perceived)
of programming distributed or cloud applications are
abstracted away, freeing the developer to focus on the particulars of
the application they are working on. In addition, the model is also
 general. It has since been shown that any computation can be
described in terms of mapping and reducing data
\cite{Zaharia2014AnAF}. Fault tolerance in this system was integrated
as an initial requirement, accomplished by utilizing recomputation to mitigate faults.
Stragglers are handled in a similar fashion, automatically recomputing
results on other nodes when results take longer than expected \cite{Dean2004}.

The most well known implementation of the MapReduce paradigm is the
Apache Hadoop project \cite{Shvachko2010}, which also includes a distributed file system.
Since its introduction, Hadoop has experienced widespread adoption by industry, government,
and academia 
as an effective means of scaling large data processing tasks to
commodity clusters and cloud instances.   

\subsubsection{TensorFlow}
TensorFlow \cite{Abadi2015TensorFlowLM} was developed and open-sourced by
Google. It was designed specifically to solve the complex task of
training deep neural networks for machine learning.
%
In TensorFlow, the basic level of abstraction for data is a tensor, an $n$-dimensional
generalization of a matrix.
These tensors pass through a dataflow
execution graph, where the vertices are computational steps and
the edges are outputs and inputs to or from these computations. This
is similar to the parameter server architecture common before
TensorFlow, but has the additional flexibility of mutable state
and concurrent execution across execution sub-graphs \cite{Abadi2016TensorFlowAS}.
A particular emphasis was placed on high level programming interface,
allowing machine learning researchers to  modify and test different
program models quickly.

%
TensorFlow can operate in distributed and heterogeneous environments.
Execution on GPUs and other specialized hardware is handled by unique
kernel implementations of the high level operation used by the
developer. When multiple nodes are being used, TensorFlow constructs
the execution graph and algorithmically places operations on specific
nodes. Once the graph is established, send and receive nodes are
placed along vertices that cross machines, to limit memory and bandwidth
when output from one vertex is transferred to two or more vertices on
another machine. These communications are generally handled by
either TCP/IP or Remote-Direct-Memory-Access (RDMA) transfers \cite{Abadi2015TensorFlowLM}.

TensorFlow has grown to prominence since its original introduction,
quickly becoming one of the de facto machine-learning frameworks.
Despite its relatively niche scope, we point it out as a successful
intersection between high level programming and performance afforded
by low level optimization.

\subsubsection{Chapel}
The Chapel programming language \cite{chapel.website} was developed by Cray with
an emphasis on productivity when developing parallel programs. Although
its syntax is similar to that of C or FORTRAN, its unique abstractions
and programming model provide a fresh take on parallel development. Chapel
is a high level, global-view parallel language with a multithreaded
execution model.

Chapel supports both data parallel and task parallel operations. For
data parallelism, the basic unit of abstraction is a \textit{domain},
or a named set of indices used to define collections for parallel
execution \cite{Chamberlain2007}. Arrays of objects are instantiated
by these domains, and Chapel supports a number of common data parallel
operations, such as \textit{forall} loops ({\em i.e.,} mappings), scans,
and reductions. Users are also free to define their own operations, as
well as fine tune the distribution of the work instead of defaulting
to  Chapel's execution strategy.

Task parallelism is achieved through \textit{cobegin} statements,
which define the beginning of a block to be executed in parallel \cite{Chamberlain2007}.
Coordination is achieved through synchronization variables and
atomic sections, which serve as a higher level alternative to
directly manipulating locks. As with data parallelism, the
exact semantics of the synchronization variables can be customized
by the developer, although this is not required.

Chapel provides many other features that are common in modern,
mainstream programming languages. Useful features such as
Object Oriented Programming and generic types come standard
in Chapel, as well as some less often seen features like
curried function calls \cite{Chamberlain2007}. 
Chapel's
integration of modern programming techniques with high level
parallelism make it a great model of joining modern programming
with classic HPC.  Nonetheless, it is not in widespread use
in classical HPC settings as of now.

\subsection{Apache Spark}
The Spark project \cite{zaharia2010spark} can reasonably be
considered to be the successor to the Apache Hadoop project. Spark has
been under active development in recent years, with a variety of
application-specific libraries distributed along with the core data
processing engine. These libraries include common routines for machine
learning \cite{Meng2015}, data stream processing \cite{Zaharia2014AnAF},
structured data queries \cite{Armbrust2015}, and graph processing \cite{Xin2013}.
It has also seen widespread adoption in industry,
and served as the basis for NADSort, a sorting system that
recently won the 2016 CloudSort Benchmark as the fastest cloud data
processing engine to sort 100TB of data \cite{nadsort.cloudsort}.

Similarly to Hadoop, Spark utilizes a MapReduce paradigm, with several
enhancements \cite{Zaharia2014AnAF}. A notable improvement is that Spark will avoid writing
intermediate results to disk whenever possible, leading to  more
efficient, in-memory execution. The fundamental layer of abstraction
in Spark is the Resilient Distributed Dataset, or RDD \cite{zaharia2010spark}. RDDs are
read-only collections of objects that can be divided into a number of
partitions, each partition with the potential to reside on a different
node in the cluster \cite{zaharia2010spark}.

When an application is executed in Spark, the driver node\footnote{The driver node
in a Spark cluster serves as the master where jobs are submitted, and high level
application flow like scheduling is determined.} will build
an execution graph of the job \cite{Zaharia2014AnAF}. There are two categories of operations
that can be performed on an RDD: transformations and actions.
Transformations are mappings of the data within an RDD, and their
execution if deferred until an action is invoked. Actions are
operations that require a usable result from an RDD, such as finding
the maximum element or printing  values to the console \cite{zaharia2010spark}.

When an action is called on an RDD, the master node will create and
potentially optimize a directed acyclic graph of the RDD's execution \cite{Zaharia2014AnAF}.
It will then schedule a number of stages, where a stage boundary is determined
by when data needs to be shuffled through the cluster. Each stage will
have one or more tasks associated with it, where a task is the basic
unit of execution. The stages and the RDD partitions are transmitted
to the worker nodes, where the tasks are executed asynchronously in
threads. Results are sent back to the master node, which schedules
further stages or passes the final result back to the application.

The Spark framework is implemented in Scala \cite{scala.language},
a derivative of Java that supports  functional and object-oriented
programming and that runs in the Java Virtual Machine (JVM). The framework also has APIs for both
the Python and R programming languages \cite{spark.website}. The communication was originally
handled by the Akka library, but since version 1.6, Spark relies on
Netty to implement TCP communication across nodes in a cluster.

\subsection{Spark on HPC Infrastructure}
As work similar to ours, we present certain efforts to integrate
Spark and its similar high level models directly into HPC
architectures. Although Spark and its family of high level frameworks
are generally designed for commodity hardware, we note that it is
possible to adapt the framework for specialized clusters and gain
performance from the unique attributes of the environment.

The Hadoop platform heavily leverages a custom distributed file system to
store results of map and reduce tasks \cite{Shvachko2010}.
The MARIANE (MApReduce Implementation Adapted for HPC
Environments) \cite{Fadika:2011:MMI:2082076.2082094} implementation
of the MapReduce model seeks to abstract
the storage component to the specific needs of the cluster. This
includes  NFS and GPFS file systems, and can be expanded to other file
systems depending on the specific cluster architecture.

In Spark, RDD's are immutable and can be recomputed from the
execution graph if a partition should be lost because  of failure \cite{Zaharia2014AnAF}.
Therefore, map tasks can be retained in memory when feasible for
a performance boost, and do not suffer from the same I/O bottleneck
of writing to disk as MapReduce and Hadoop. However, Spark will experience a
similar performance degradation when map results spill over to disk and
shuffling partitions across the cluster. Chaimov et al.\@ 
noted that repeated reads and writes to common files in Spark causes
metadata operations that are significantly more pronounced on HPC
distributed memory systems than commodity hardware with
local storage \cite{Chaimov2016ScalingSO}. They investigated
techniques for removing this and similar bottlenecks when deploying
Spark on the Lustre file system and observed effective scaling
of up to $O(10^4)$ cores in an HPC installation.

Another consideration of running Spark on  an HPC infrastructure is the
scheduler and cluster manager. Spark can be deployed manually on a
commodity cluster, or through a dedicated resource manager like
Apache Mesos \cite{Hindman2011} or Apache YARN \cite{KumarVavilapalli2013}.
This can
create a compatibility issue when attempting to run Spark as middleware
on existing installations with their own schedulers and resource managers.
Baer et al.\@ successfully implemented an integration of Spark into the
PBS batch scheduling environment \cite{Baer2015}.
This and the other
integrations mentioned above demonstrate that Spark, while designed for commodity hardware
clusters, can and has been modified to  operate on
high performance infrastructure correctly.

\section{Approach}
\label{sec:approach}
Inspired by the history of MPI and the power and popularity of Apache
Spark, we sought to implement an effective, peer-to-peer message passing scheme in
Spark to demonstrate  potential benefits. We note that, while
similar in end goal, our approach integrates a core facet of HPC into
the high level framework itself, rather than running Spark ``as-is'' or nearly so in
an HPC environment, or attempting to expand MPI to emulate a
MapReduce model. 

\subsection{Communication Implementation}
The MPIgnite framework introduced here serves as a prototype for integrating traditional
HPC concepts in higher level environments, with the specific instance
being message passing inside Apache Spark. For the purposes of this
work, we sought to leverage  existing communication infrastructure
already in the Spark code, leading to a relatively simple modifications of
the original project. This demonstrates the feasibility of combining the
two paradigms in practice while also leaving room for specialized, optimized
implementations when and where appropriate.

As with all distributed systems, communication between nodes in Spark is
fundamental. The general architecture of Spark is a master-worker model.
Even when Spark is executed locally on a single machine, tasks
are transmitted to worker threads, and results and communicated back
to the master. In the version of Spark we derived our work from (2.1), these
communications are implemented as RPCs on top of asynchronous Scala futures
in a local deployment and via TCP/IP connections in a cluster.

Since the groundwork for a message passing scheme was already built into
Scala, we sought to repurpose it to suit our needs. Therefore, we introduced our
own RPC methods specific to sending and receiving messages. We additionally
created an API so that developers could access these communication methods
in their driver programs with signatures similar to that of traditional
MPI (discussed below). Since Scala natively supports futures and promises
\cite{scala.language}, implementing non-blocking versions of both sending
and receiving functions was straightforward.

 Spark abstracts communication through RPC ``endpoints'' internally, which
are interfaced through \texttt{RpcEndpointRef} reference objects. A single
endpoint can have multiple references, and any reference can communicate
through the endpoint. In both a local and clustered deployment, the workers
have established endpoints to communicate with the driver. These
communication channels are typically used to transmit status information,
such as when a task completes or unexpectedly fails. For our purposes, we
utilize these same endpoints to send and receive generic application
messages.

For local deployments, the previous description is sufficient, since there
is only one worker node. More care is needed for proper peer-to-peer
message passing in a clustered deployment. In that case, scheduled tasks
are distributed along with a mapping of the process rank to the unique
worker identifier that is executing that process. When a message is being
sent to another process, the worker will check the mapping to see if it
has an RPC endpoint associated with that worker. If it does not, it requests
the addressing information of that worker to establish an endpoint
before transmitting the message.
Workers maintain a collection of RPC endpoints for workers that gets augmented
 on an as-needed basis. This amortizes the cost of sending to new worker
nodes as well as potentially reducing the number of networking connections
for a given  topology. Additionally, we buffer messages on the
receiving worker, meaning that no network communication is necessary for
receiving a previously sent message\footnote{We recognize that this type of 
buffering is quite different from what typical MPI implementations choose to do. The first goal is functionality; we will work on enhanced scalability and performance in subsequent releases.}.

Implementing MPIgnite's communicators was facilitated by Scala's object-oriented
paradigm, since they can be naturally be utilized as objects with communication
methods. In our framework, each communicator object maintains a mapping of the
ranks going from the rank within the commmunicator to the rank in the default,
or world, communicator. When a communicator is split to create a sub-communicator,
every process participating in the split sends a message of its global rank, key
and color to the lowest process rank participating in the split. That root process
receives all the split information, groups it by color, and sorts it according to
key. The sorted data is then configured to be a new rank mapping before broadcast
back to the group.

To ensure that messages are not passed across subcommunicators, each communicator
contains a context identifier field that is specific to the group that participated
in the aforementioned split. Messages sent from that communicator are passed
along with that identifier, and checked for equality at the receiving end to
ensure can only occur within similar communicators. To facilitate splitting,
the global communicator always has an identifier of 0, so internally messages
can be sent and received directly.

In our initial implementation of MPIgnite, all communications passed through
the master node. Subsequent iterations advanced the model to allow for
actual peer-to-peer communication by permitting messages to be sent
directly between worker nodes.
These two implementation iterations naturally gave rise to an interesting approach to
the fault tolerance problem. Although not investigated in this work, we can potentially
switch between peer-to-peer mode and master-worker mode internally when coping with faults.
After recovery, peer-to-peer 
communication would resume. 

\subsection{Parallel Closures}
The most important facet of our design is the API directly available to
developers for programming their parallel code. We wanted an interface
that would be both familiar to long-time MPI users as well as approachable
to Spark uses who may have never written explicitly parallel code 
before. In addition, we wanted our environment to be flexible, allowing
the user the freedom to use whichever style of programming  best
fit the domain of the application.

Our result from these goals is a system by which parallel sections
of code are written as function closures. Since Scala is a functional
language that supports functions as first-class objects, this was a
logical solution that proved natural to implement. This is similar to
Chapel's method of defining parallel blocks with \textit{cobegin}
statements, except that our model uses full-fledged functions that
can be referenced, imported, and passed as arguments.
These functions can be defined anonymously at the location they are used,
or defined elsewhere and be reused, leading to a more modular design than
Chapel's parallel blocks.

Once a parallel function has been defined (either anonymously or as a
named value), the developer passes it to a \texttt{parallelizeFunc} method
that MPIgnite provides to create an RDD (analogous to Spark's
\texttt{parallelize} function for creating an RDD
from a dataset). From there, the user can
call \texttt{execute} on the RDD to initiate the parallel execution.
The number of threads of execution can be selected at runtime by
a parameter passed to the \texttt{execute} function.
The result of the execution will be an array of  return values from
each process, if any. This is nearly identical to the process of 
creating and transforming a typical RDD in standard Spark. 
Developers can select which portions of their applications should 
be executed in parallel by how they construct and utilize their 
parallel functions. Moreover, the process is interoperable with 
traditional data-parallel RDDs, enabling a single application to
realize task (and/or data) parallel message passing without foregoing 
  off-the-shelf Spark.

The closures can be as long or short as needed. Longer closures
will prove more scalable, since the end of a closure forms an
implicit synchronization barrier within the driver application.
That is, once a closure is executed in the driver application,
all instances of the parallel function must complete before the
driver program can continue. Future work will explore potentially
chaining these closures together asynchronously, affording large
parallel sections composed of several smaller ones.

 Parallel closures in our model do
not currently support arguments other than the required
\texttt{SparkComm} object (discussed below). We did not see the
apparent need for this, since these closures have access to
variables in their outer scope, demonstrated below in
an example. Furthermore, parameters can be achieved by wrapping
the closure in another function that does accept any  needed
parameters.

\subsection{Similarities to MPI}
Since one of our main goals was to develop an environment that would
be familiar to long-time MPI developers, we made a number of design
decisions to remain ``in the spirit'' of the standard. However, we specifically
chose not to be backwards compatible or fully implement the function
signatures of the standard for a number of reasons. Our main motivation
for breaking from the standard was to bring a fresh perspective to an
established programming model. Additionally, implementing all or even
most of the standard would be technically substantial, particularly in
the narrow scope of our prototype. Additionally, many language
features leveraged freely in the C and FORTRAN implementations of MPI
standard (such as out parameters)
are not feasible within the Scala language.

Instead of attempting to reimplement MPI inside of Spark, we sought a
blending (compromise of sorts) of the two models that would complement the features of
both. As discussed in the previous section, parallel sections of an
application are crafted inside of functions that are transformed into
RDDs. Each function executed in parallel can be thought of a single
instance of a true MPI program. Similarly, each instance of the parallel
execution has a unique rank, and the total number of instances can be
collected at runtime. These values are gathered from the \texttt{SparkComm}
object, which is passed in to every instance of the function.

The \texttt{SparkComm} object in a parallel block is also used to
send and receive messages, much like a communicator in MPI.
Processes are denoted by their unique rank number
and messages can be tagged for proper identification, as in MPI. Group
communication is implemented from these primitives, though a possibly
more efficient strategy is to utilize Spark's built-in broadcasting
\cite{Chowdhury2010} (to be considered as future work).

\texttt{SparkComm} objects can also be split into subcommunicators,
in almost identical fashion to MPI. Our prototype also features
the collective operations \texttt{broadcast} and \texttt{allReduce}.
These are explained by example in the proceeding section. A notable enhancement of
the \texttt{allReduce} function is that MPIgnite supports passing arbitrary
reduction functions, fostered by the functional nature of the Scala language.

\subsection{Communication Data Types}
The general syntax of sending and receiving messages is similar to MPI,
though with key differences. The \texttt{SparkComm} object
defines \texttt{send} and \texttt{receive} methods for communication.
However, instead of sending and receiving
data buffers, true Scala objects make up the content of messages,
provided those objects are serializable.
This creates a simple, object oriented interface appropriate for Scala-style
programming. In addition, calling
\texttt{receive} evaluates to the object that is received in the blocking
case, or a future of the object in the nonblocking case.
The ability to send and receive first-class objects permits a higher level
of abstraction when using our model. Data structures do not need to be derived
or to have message structures that are tightly coupled to their implementation.


\section{Examples}
\label{sec:examples}

The following examples provide a brief hands-on introduction to
the usage of the MPIgnite framework. Emphasis has been placed
on developing and executing the parallel blocks, and also the
interprocess communication interfaces, all while keeping the
samples small and easily digestible\footnote{These and other
examples will be made available for download via GitHub and include
instructions for installation and execution.}.

The first listing below depicts a simple example of matrix-vector multiplication, 
and does not use any explicit interprocess
communication. We start by defining the matrix and vector as
two-dimensional and one-dimensional arrays, respectively (in Scala,
arrays are created and indexed by parentheses). Next we define our
parallel code block as both an anonymous function and an argument to
the \texttt{parallelizeFunc} method on the \texttt{SparkContext} object
available to all Spark applications (denoted in the examples as the
\texttt{sc} variable). The square brackets after
the \texttt{parallelizeFunc} name indicate a type parameter of
integer, since each parallel instance of our function will return
a single integer of the row-vector multiplication. We then acquire
the current process's rank through the \texttt{SparkComm} object
to determine if this process is needed for any valuable work. Then we
compute and return the result, the return value implicitly are the
last expression of a function. Once the closure is parallelized, we
\texttt{execute} it with eight concurrent instances and sum the
partial results in the driver application.

\begin{lstlisting}[language=Java, caption=Matrix Multiplication with MPIgnite]
val mat: Array[Array[Int]] =
  Array(
    Array(1, 2, 3),
    Array(4, 5, 6),
    Array(7, 8, 9)
  )
val vec = Array(1, 2, 3)

val res = sc.parallelizeFunc[Int](
    (world: SparkComm) => {
  val rank = world.getRank

  if (rank < mat.length) {
     mat(rank).zip(vec)
       .map(a => a._1 * a._2).sum
  } else 0
}).execute(8).sum
\end{lstlisting}

{\sloppypar This example could have equivalently
been written with traditional RDDs and a mapping
function. However, being able to program in a task
parallel setting is a good fit on to the problem domain,
and does not sacrifice existing options. Additionally,
since the scope of parallel sections is limited to
function blocks, developers can incorporate both
task parallel sections or traditional RDDs
depending on their programming preferences and the application domain.}

{\sloppypar The next example depicts a simple ring application to show the
message passing API of MPIgnite. In this instance, we define a function
\texttt{ring} explicitly before parallelizing it.
Similar to the previous example, the rank and size of the process
is collected from the \texttt{SparkComm} object. Each process
sends the token variable to the process after it, with the root
process starting and receiving the token from the last process.
Since \texttt{receive} in this example is blocking, no process
other than the root will send until it has received the token.}

The \texttt{send} function takes three arguments: the rank of
the process to which data is being sent, the tag associated with
the message, and the data object that is to be transmitted.
The \texttt{receive} function has two arguments for the rank
of the sending process and the message tag. Receiving also requires
an additional type parameter (denoted by the square brackets) to
indicate the type of data that is to be received. This is necessary
to permit proper deserialization and casting.

\newpage 
\begin{lstlisting}[language=Java, caption=Message Passing in MPIgnite]
def ring(world: SparkComm) = {
  var token = 0
  val rank = world.getRank
  val size = world.getSize
  
  if (rank == 0) {
    token = 42
    world.send(rank + 1, 0, rank)
    token = world.receive[Int](size - 1, 0)
  } else {
    token = world.receive[Int](rank - 1, 0)
    world.send((rank+1) % size, 0, token)
  }
}

val parallel = sc.parallelizeFunc(ring _)
parallel.execute(16)
\end{lstlisting}




To demonstrate nonblocking receive (recall that sending in MPIgnite is always
nonblocking), our next example uses it exclusively. Making a nonblocking call
to receive returns a future object, or a read-only placeholder for an asynchronous
computation. Futures can be explicitly waited on or can have callbacks defined to
execute on their success for failure.

\begin{lstlisting}[language=Java, caption=Nonblocking receive example, upquote=true, showstringspaces=false]
import scala.concurrent._
import ExecutionContext.Implicits.global

def evenOrOdd(sc: SparkContext) =
  sc.parallelizeFunc((world: SparkComm)=>{
    val (size, rank) =
      (world.getSize, world.getRank)
      
    if (rank < 5) {
      world.send(rank + 5, 0, rank)
      val f =
        world.receiveAsync[Boolean](
          rank + 5, 0)
      println(s"Rank $rank: Waiting...")
      f.onSuccess {
        case b =>
          println(s"$rank is even: $b")
        } 
      } else {
        val r = world.receive[Int](
          rank - 5, 0)
        Thread.sleep(3000)
        world.send(rank - 5, 0,
          r % 2 == 0)
      }
  }).execute(10)
\end{lstlisting}

For the simplicity of the example, we initially import
the \texttt{ExecutionContext.Implicits.global}. In Scala,
futures are executed asynchronously through some
\texttt{ExecutionContext} that can be explicitly created
as a thread pool and passed to the \texttt{receiveAsync}
function. In the above example, using the import allows
the futures to implicitly utilize a global static thread pool.
This is occasionally unadvisable, but convenient for our
example.

Futures in the Scala language can be utilized by
registering callback functions to execute with the result
of the asynchronous computation upon completion, as done
in the previous example. Alternatively,
futures can be explicitly waited on by using the \texttt{Await.result(f)}
function with the argument the future in question. This method of
synchronization is analogous to the \texttt{MPI\_Wait} function.

As a more complete example of the features of MPIgnite,
the next listing extends the previous matrix-vector
multiplication to include a 2D decomposition.

\begin{lstlisting}[language=Java, caption=Matrix-vector multiplication with 2D data decomposition]
sc.parallelizeFunc((world: SparkComm) => {
  val worldRank = world.getRank
  val row = world.split(worldRank / 3,
    worldRank)
  val col = world.split(worldRank % 3,
    worldRank)
  val a = worldRank + 1
  val rowRank = row.getRank
  val colRank = col.getRank
  
  // Distribute the vector to the diagonal
  if (rowRank == row.getSize - 1)
    row.send(col.getRank, 0, 1+col.getRank)
  val x_row = if (rowRank == colRank)
    Some(row.receive[Int](
      row.getSize - 1, 0))
    else None
  val multiplied = x_row match {
    case Some(x) =>
      col.broadcast[Int](colRank, x)
      x * a
    case None =>
      a * col.broadcast[Int](rowRank)
  }
  val result = row.allReduce[Int](
    multiplied, (a: Int, b: Int) => a + b)
}).execute(9)
\end{lstlisting}

This example is specifically tailored to the 3x3 case presented
earlier, but similar decompositions can be formed for non-square
matrices of arbitrary size, since MPIgnite has the basic communication functions.
This example also employs several Scala language features like
pattern matching and Options. We should note that recipients of
a \texttt{broadcast} message only need to indicate the root rank
of the broadcast.

For completeness, Figure~\ref{fig:api-comparison} lists the MPIgnite
methods and the corresponding MPI functions. More methods will
be augmented in future work. 

\begin{figure*}[ht]
\centering
\begin{center}
 \begin{tabular}{|c | c|} 
 \hline
 {\bf MPIgnite} & {\bf MPI} \\
 \hline \hline
 comm.send(rec, tag, data) & MPI\_Send \\ 
 \hline
 comm.receive[T](sender, tag): T & MPI\_Recv \\
 \hline
 comm.receiveAsync[T](sender, tag): Future[T] & MPI\_Irecv \\
 \hline
 Await.result(f: Future[T]): T & MPI\_Wait \\
 \hline
 comm.getRank & MPI\_Comm\_rank \\
 \hline
 comm.getSize & MPI\_Comm\_size \\
 \hline
 comm.split(color, key): SparkComm & MPI\_Comm\_split \\
 \hline
 comm.broadcast[T](root, data\footnote{Only requried for root process}): T & MPI\_Bcast \\ [1ex] 
 \hline
 comm.allReduce[T](data, f(a, b): T): T & MPI\_Allreduce \\ [1ex] 
 \hline
\end{tabular}
\end{center}
\label{fig:api-comparison}
\caption{Comparison of the MPIgnite and MPI function signatures}
\end{figure*}

\section{Discussion}
\label{sec:results}


Apache Spark has experienced broad success in recent years, both as
an open source project and as an efficient data processing
engine. Our work contributes to that success by incorporating a
message passing scheme, substantially enhancing the programmability
of the environment. This not only enables for established
Spark developers to utilize task parallelism explicitly as they
see fit, but also benefits authors of existing HPC applications by
giving them access to a high level framework without having to
compromise their core MPI algorithms.

The closure syntax for developing parallel code sections
was easily made possible by the high level nature of the
Scala language. This pattern for developing parallel applications
allows the parallel portions to be reasoned about and developed
independently, permitting greater modularity and enhanced ease of
maintenance. This in itself is not entirely new, since other
languages like Chapel permit independent parallel blocks, but our
framework does so through first-class functions. This means that
they are be easily extended, wrapped, passed as arguments, and
reused. Entire libraries can be written of common parallel
functionality and serve as building blocks for complex parallel
applications.

Furthermore, our work does not compromise the integrity of the
Spark platform. A single application can support both parallelized
functions unique to MPIgnite as well as typical RDDs found in any
Spark application. As such, the plethora of Spark libraries and
resources remain pertinent to MPIgnite developers via this
backwards compatibility. This combination of specialized data
parallelism with message passing task parallelism permits users of
both camps to cross sides as they see fit. As parallel
sections are independent and narrow instead of global as in MPI,
they can harmoniously coexist when the problem domain requires a
diverse strategy.

Despite the benefits previously described, MPIgnite will obviously
not be the best tool for every job. In spaces
where latency is absolutely critical, a high level approach such
as ours will likely be unsuitable without considerable optimization effort.
However, we note that our platform can still benefit those
use cases, albeit indirectly. Since Scala (and thus MPIgnite) is
so high level, it provides an environment that can be used
for quick experimentation and rapid development. Therefore, MPIgnite can
serve as a prototyping tool for testing the feasibility of
approaches and algorithms to problems before seeking a 
potentially costlier bare-metal solution.  

\section{Future Work}
\label{sec:future-work}
The MPIgnite platform in its current form serves as a prototype to
demonstrate the potential benefits of integrating HPC concepts into
high level programming frameworks common in cloud computing.
As such, there are many interesting avenues to extend this work
in pursuit of its initial goals. The most obvious extension
would be to create a
more efficient implementation of the message passing scheme than
presented here. Although a Scala-based JVM implementation will
likely never reach the speed of a bare-metal C-based distribution of
MPI, competitive performance would make the MPIgnite model even
more attractive in addition to the reasons already discussed.

In adition to efficiency, we hope to explore MPIgnite's potential
for scalability. Since MPIgnite was
created with strategic additions to the core Spark framework,
it stands to reason that MPIgnite should scale as well as Spark
does. Additionally, techniques for scaling like creating a
hierarchy of master nodes to coordinate potentially thousands
of processes could be feasible with augmentations to Spark's
scheduling system.

Another area of future work that we are planning is a proper
analysis of the closure model for parallel programming. Since
Scala is very high level and natively supports these anonymous
functions, they were a natural choice for our implementation of
MPIgnite. This feature is not available in C or FORTRAN, so
investigating this model and the potential to build reusable
libraries of common and extensible parallel operations would
be a worthwhile contribution to making the original MPI model
more dynamic and extensible.

Also, providing sufficiently efficient 
adapters  to incorporate  3rd-party MPI libraries (e.g., math libraries)
 is an important area of future work.
Allowing these libraries to be instantiated with sufficient MPI support, or using
 a traditional MPI middleware underneath, yet enabling the present framework
 to function mostly unhampered, will require significant design work, experimentation,
 and engineering effort.

Finally, although briefly mentioned in this communication, this
work can be extended with a thorough examination of integrating
Spark's fault tolerance with the message passing scheme. Fault
tolerant MPI is an active area of research, and our model
presented here provides a unique approach by building MPI into
a framework that enables fault tolerance.

\section{Conclusion}
\label{sec:conclusion}

We chose the Apache Spark platform for our work in large part
because of  its popularity in the sphere of high level cloud 
computing. We sought to maximize impact by combining that general-purpose computing platform
with the de facto standard of HPC, MPI, with one of
the largest and most general-purpose high level computing
platforms. The result is a framework that will ``feel familiar''
to long-time MPI developers, without compromising the power
and generality of Spark itself. This opens the door for
cloud computing developers to utilize the well studied algorithms
and techniques of HPC in the comfort of their own framework.
Similarly, traditional HPC developers can employ high level
language concepts and sophisticated data parallelism for easier
development and rapid prototyping.


\section*{Acknowledgements}
This material is based upon work supported by the National Science
Foundation under Grants Nos.\@ 1562659 and 1229282.  Any opinions, findings, and
conclusions or recommendations expressed in this material are those of
the authors and do not necessarily reflect the views of the National
Science Foundation.

We acknowledge the previous work and contributions of Mr.\@ Jared Ramsey in his MS thesis at Auburn that motivated this work.
Dr.\@ Jonathan Dursi's blog \cite{hpc.dying.blog} was a strong motivator for this work.

Dr.\@ Purushotham Bangalore provided helpful input to this paper.

\nocite{DBLP:journals/concurrency/SkjellumWLWBLSM01,hpc.dying.blog}
\bibliographystyle{abbrv}
\bibliography{mpi-plus-spark}


\end{document}